# Thermodynamic of Glass Indentation: Dissipative Aspects.

Guglielmo Macrelli

Isoclima SpA. R&D Department

e-mail: guglielmomacrelli@hotmail.com

**Key words**: Glass, indentation, dissipative effects, thermodynamics

**Abstract**:

A criterium is derived to understand the relevance of thermal effects resulting from high rate mechanical actions on glass surface. The criterium is based on the concept of characteristic contact time of the load to the glass surface. This criterium is of particular relevance to impact phenomena. A general discussion about dissipative aspects of indentation on glass surface is presented and a thermodynamic approach is suggested to evaluate the energetic of the process. Further generalization is proposed based on the thermodynamics of continuous media.

## I - Introduction

The macroscopic mechanical behavior of glass indicates a fully elastic characteristic up to the breakage which occurs without any previous indication of the upcoming failure. This breakage mode is frequently addressed as "brittle fracture" [1]. Looking to the microscopic and nanoscale behavior of glass there is clear evidence of non-elastic phenomena [2]. Both densification and shear effects indicate that, at nanoscale, glass do not behaves elastically exhibiting the presence of dissipative phenomena. The effects of mechanical actions on glass surfaces is also depending on the load rate. As demonstrated by Lawn et al [3] high-rate load experiments (as high-speed impacts) may lead to a significant temperature increase leading to glass melting. A criterium is derived here to better understand the relevance of thermal effects resulting from load rate. A further generalization of the thermodynamics of indentation is proposed based on the concepts of mechanics of continuous media.

## II – Indentation as a dissipative process.

As pointed out by Varshneya et al [3] and Lawn et al [4] the stress field generated by an indenting body is elastic-plastic in nature. Lawn et al. assume that, because of the plastic component, part of the indentation work must be dissipated in the contact area as heat. Additionally, assuming high indentation rates, they approximate the heat generation as an adiabatic process. This statement can be generalized assuming that the dissipative components of the indentation work are due to the affinities (increase of entropy of the system by changing the set of the thermodynamic coordinates). The indentation is a cyclic process where a load $F$ (force) is applied to a material at its surface on a localized point, up to a maximum value $Fm$ than released down to zero. The response of the material can be characterized by the amount of deformation $h$ that will increase to a maximum value $hm$ in correspondence with $Fm$ and will decrease down to a value $hr$ as $F$ will be totally released. In a fully reversible process, the unloading Force/displacement curve shall follow the loading curve returning to $hr$ equal to zero. A final residual deflection $hr \neq 0$ indicates a hysteresis effect (see figure 1). The appearance of a hysteresis is an indication of the thermodynamic irreversibility of the cyclic indentation process.

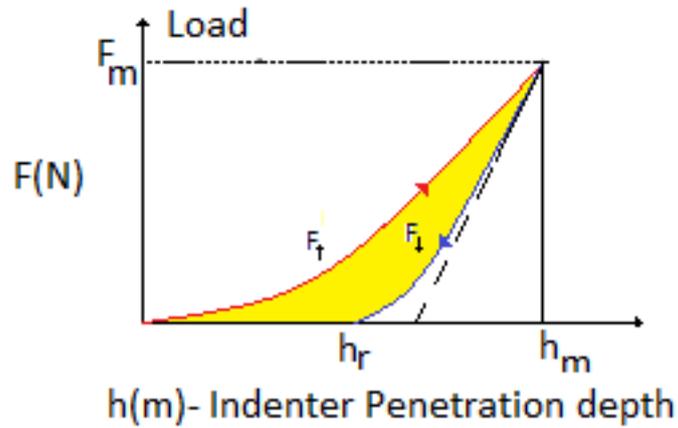

Figure 1 – Indentation cyclic process diagram.

The mechanical work during the loading $W_\uparrow$ and unloading $W_\downarrow$ phases can be expressed as:

$$W_\uparrow = \int_0^{hm} F_\uparrow(h)\,dh \tag{1}$$

In the loading phase, and:

$$W_\downarrow = \int_{hm}^{hr} F_\downarrow(h)\,dh \tag{2}$$

during unloading phase. The total work exerted by the indenter in the cyclic indentation is just the sum of the two parts:

$$Wt = W_\uparrow + W_\downarrow = \int_0^{hm} F_\uparrow(h)\,dh + \int_{hm}^{hr} F_\downarrow(h)\,dh = \int_0^{hm} F_\uparrow(h)\,dh - \int_{hr}^{hm} F_\downarrow(h)\,dh \tag{3}$$

In case of reversible response of the material, $Wt=0$, while the evidence of $Wt > 0$ indicates the presence of dissipative phenomena occurring in the material during the loading and unloading of the indentation. Looking at figure 1 it appears that $Wt$ is represented by the area between the loading and unloading curves. The ratio:

$$We = \frac{\int_{hr}^{hm} F_\downarrow(h)\,dh}{\int_0^{hm} F_\uparrow(h)\,dh} \tag{4}$$

It is called the elastic recovery of the material, $We=1$ (100%) when the material response is reversible (elastic behavior) $We < 1$ (< 100%) when we have plastic, dissipative components in the material response.

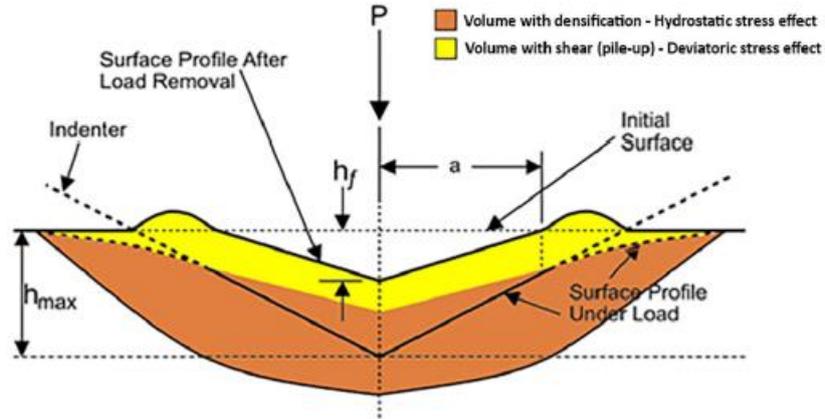

Figure 2 Visualization of Vickers indent cross section after load retraction. Yellow zone is the shear plasticity area with edge pile-up, muddy orange zone is the densification zone. From Varshneya et al. [3]

## III. Thermodynamic of indentation

Let's consider the indentation process on glass as a thermodynamic system whose boundaries are encompassing the portion of glass affected by inelastic phenomena (see figure 3). The indentation may be considered as an external mechanical work performed on the thermodynamic system.

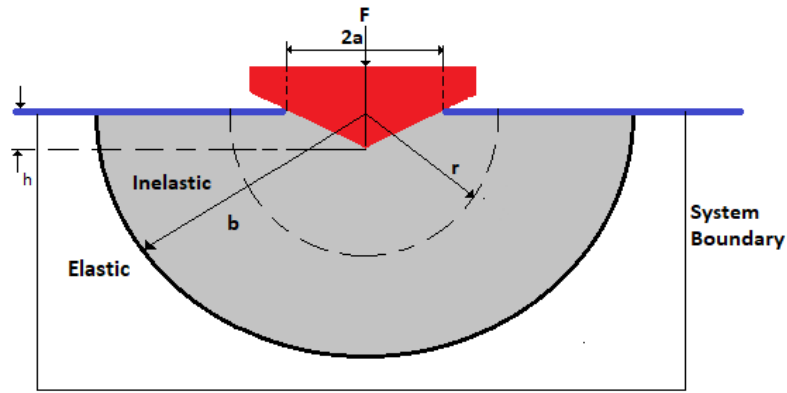

Figure 3 – Schematic representation of the system considered for thermodynamic analysis.

The thermodynamic laws written for the system of figure 3 read [5]:

$$\frac{d(U+K)}{dt} = \dot{W} + \dot{Q} \qquad (5)$$

$$\frac{dS}{dt} = \Phi_s + \Pi_s \qquad (6)$$

Where :

*U (J)* – Internal energy

*K (J)* – Kinetic + rotational energy

*W (J)* – Mechanical work on the system

*Q (J)* – Heat transferred to the system

*S (J/K)* – Entropy

*Φ_s (J/sK)* - Entropy flux rate from the system boundaries

*Π_s (J/sK)* – Entropy source rate within the system

From the system definition (The system is constrained so no translational nor rotational component are considered and *K=0*) we can exclude heat transfer to the system and, at the same time, entropy flux from the system boundary. The only entropy fluxes could occur through the boundary Elastic/Inelastic, this means that, choosing a suitable overall boundary that encompass the interface surface Elastic/Inelastic, we can exclude entropy fluxes in or out of the "system". Hence, Thermodynamic equations result:

$$\frac{dU}{dt} = \dot{W} \; ; \; \frac{dS}{dt} = \Pi_s \geq 0 \tag{7}$$

The first equation is nothing but the one considered in the Lawn et al. paper. In their paper the only entropy source they considered is the energy dissipation in heat. Because of the adiabatic hypothesis they assume this goes all into a temperature increase:

$$f\delta W = m_b \cdot c_p . \Delta T = \frac{\pi b^3}{3} \cdot \rho \cdot c_p \cdot \Delta T \tag{8}$$

Where *fdW* is the dissipated fraction of the work (*Wt*) performed during the indentation cycle (the area in between the loading and unloading curves of Figure 1) $m_b$ *(Kg)* is the glass mass below the indenter and within the inelastic semi sphere of radius *b (m)* (see figure 2), $\rho$ (kg/m$^3$) is the glass density and *cp* (J/kgK).is the specific heat of glass. Please note that the equation herewith is different from the one of Lawn et al. because here we consider a semi sphere of glass of radius *b* instead of a sphere. In the Lawn et al. analysis there are two approximations taken as assumptions:

A) All dissipated work (energy) is transformed into heat
B) The system is adiabatic

In the entropy source rate term, there should be included a number of additional contributions (see for example Boley – Weiner [6]), for the: hydrostatic and deviatoric components of the stress tensor for densification and viscous effects, mass transfer terms related to stress gradients and fracture energy terms for cracks nucleation at the inelastic/elastic boundary and further development of fractures. The second assumption leads to the values of temperature increase reported by Lawn et al. As correctly pointed out by Lawn et al. the adiabatic assumption is valid as far as the contact time of the indenter is very limited (they speak of milliseconds) which is compatible with high-speed impacts. In the following section a quantitative criterium is derived to better evaluate the limits of assumption B).

**IV. A criterium for adiabatic hypothesis and a criterium to establish the "yield effects" size.**

We can develop here a simple criterion to establish the time below which we could expect the validity of the adiabatic hypothesis. In doing that we consider that the heat generated in the indentation would produce a heat flux through the boundary driven by a temperature gradient $\Delta T/b$. The energy $E_b$ (J) transferred through the boundary can be estimated:

$$E_b = \lambda \frac{A}{b} \Delta T \cdot \tau \tag{9}$$

Where $A$ (m$^2$) is the area of the surface of the interface Inelastic/Elastic semi sphere, $\lambda$ (W/mK) is the glass thermal conductivity and $t$ (s) is the time. The characteristic time for heat transfer from the zone where it is generated to the rest of the body $\tau$ can be estimated by equating $E_b$ with $fdW$, hence we get:

$$\tau = \frac{\rho \cdot c_p}{3\lambda} \cdot b^2 \tag{10}$$

The adiabatic condition assumed by Lawn et al [4] is valid if the load contact time $tc$ is far grater than the characteristic time $t$. So the criterium reads:

$$t_c << \tau = \frac{\rho \cdot c_p}{3\lambda} \cdot b^2 \tag{11}$$

Assuming a value of $b$ (limit of Inelastic/elastic zone) of the order of 10$^{-3}$ m and typical values for thermal conductivity $\lambda = 1$ W/mK and specific heat $c_p = 800$ J/kgK for silicate glass, the characteristic time results $\tau = 0.5$s. This means that, if the contact time $tc << \tau$ than adiabatic conditions are met while, if $tc \approx \tau$ we can expect a significant temperature reduction in respect to the one reported by Lawn et al.[4] for fully adiabatic conditions. The conclusion is that, in quasi-static indentation with loading and unloading cycle of the order of magnitude of seconds it is very unlikely that temperature exceed the glass transition temperature while for impact phenomena with a contact time of milliseconds the adiabatic assumption of Lawn et al. [4] is justified.

A relevant criterium to characterize the size where yield effects become relevant can be established following approaches already suggested in the literature by Dugdale [7] for metals and further discussed by Barenblatt [8] for brittle fracture. Simply considering the stress at yield $\sigma_y$ and the critical stress intensity factor $K_{IC}$ the dimension of yield effects area can be estimated:

$$R = \frac{\pi}{8} \left( \frac{K_{IC}}{\sigma_y} \right)^2 \tag{12}$$

Yield stress for silica glass has to be at least equal to the maximum measured tensile strength of glass [1] $\sigma_y = 13$ GPa. Substituting 0.8 MPa m$^{1/2}$ for $K_{IC}$ [1], the estimated plastic zone size equals 1.5 nm, which is several times the size (0.5 nm) of the silica tetrahedra rings in silica glass. This last criterium clearly indicates that inelasticity effects are relevant at nanometric scale while at macroscopic scale the hypothesis of elastic brittle material is quite acceptable.

**V Discussion**

From a continuum mechanics perspective [9] the stress field resulting from a mechanical action on glass can be separated in spherical (hydrostatic) and deviatoric components generating, respectively, densification and shear pile-up effects. Glass yielding at nanoscale causes permanent changes in dimension or shape upon stress application. These changes are: permanent volume change due to densification – hydrostatic / spherical stress component while permanent shape change are caused by shear yield (Sliding/Twisting) to be addressed to the deviatoric stress component.

$$\sigma_{ij} = \frac{\sigma_{kk}}{3}\delta_{ij} + \sigma_{ij}^D = \sigma_H \delta_{ij} + \sigma_{ij}^D \qquad (13)$$

The Strain Energy $W$ is the scalar product of stress $\sigma$ and strain tensors $\varepsilon$ and it can be separated in the spherical and deviatoric components as follows:

$$W = \sigma : \varepsilon = \sigma_H : \varepsilon_H + \sigma^D : \varepsilon^D \qquad (14)$$

The analyisis of dissipative effects can be generalized :through a thermodynamic approach based on the Clausius-Duhem Equation

$$\rho(\dot{u} - T\dot{s}) - \sigma : d \leq -\frac{\vec{q}}{T}\vec{\nabla}T \qquad (15)$$

Where $u$ and $s$ are respectively internal energy and entropy per unit mass and $d$ is the deformation rate tensor and $q$ is the heat flux transferred to the system and $T$ is the absolute temperature and $d$ is the strain rate (rate of deformation) tensor. Further studies are needed to apply equations (13), (14) and (15) in order to come to workable solutions clarifying the dissipative effects of glass surface indentation through a systematic thermodynamic approach.

**VI – Conclusion**

Irreversible dissipation effects shall be finally converted to heat. Evidence of densification, shear effects and inelasticity indicate that, at a nanoscale, glass failure (that is fracture) is anticipated by "yield" phenomena that, apparently, are not evident at a macroscopic scale. The plasticity zone size, where inelastic yield effects generate, can be estimated from continuum Dugdale–Barenblatt criterium (12) of a plastic zone at a crack tip. A thermodynamic approach is proposed take into account dissipation effects. A criterium (11) is derived to evaluate the adiabaticity assumption in high rate (impact) mechanical actions on glass. A systematic general thermodynamic framework is proposed, based on continuum mechanics, to generate a systematic thermodynamic theory of indentation on glass surfaces.

**References**

[1] E.Le Bourhis, Glass, Mechanics and Technology, 2nd Edition, Wiley-VCH, Weinheim Germany, 2014.

[2] T.Rouxel, J.I.Jang, U.Ramamurty, Indentation of glass, Prog Mater Sci, 2021;121:100834

[3] A.K.Varshneya, G.Macrelli, S.Yoshida, S.H.Kim, A.L.Ogrinc, J.C.Mauro Indentation and abrasion in glass products: lessons learned and yet to be learned, Int J Appl Glass Sci, 2022;13:308-337https://doi.org/10.1111/ijag.16549

[4] B.R. Lawn, J. Hockey, S.M. Wiederhorn, Thermal effects in sharp-particle contact; J Amer Ceram Soc, 1980;63: 356-358]

[5] J.P.Ansermett, S.D. Brechet, Principles of Thermodynamics, Cambridge University Press, Cambridge UK, 2019

[6] B.A. Boley, J.H.Weiner, Theory of Thermal Stress, John Wiley and Sons, New York USA,1988

[7] G.S. Dugdale, Yielding of steel sheets containing slits, J Mech Phys Sol, 1960;8:100-104

[8] G.I.Barenblatt, The mathematical theory of equilibrium cracks in brittle fracture, in H. L Dryden, T. Von Harman (eds), Advances in Applied Machanics. Vol. 7. New York, Academic Press; 1962. pp. 55-129.

[9] J.Botsis, M.Deville, Mechanics of Continuous Media: an Introduction, EPFL Press, Lausanne CH, 2018.